\documentclass[letter,twocolumn]{jpsj3}
\usepackage{txfonts}
\usepackage{dcolumn}
\usepackage{color}
\usepackage{bm}
\usepackage{nicefrac}
\usepackage{amssymb}

\title{ Anisotropy of the Upper Critical Field in the Heavy-Fermion Superconductor UTe$_2$ under Pressure}

\author{Georg \textsc{Knebel}$^{1}$,  Motoi \textsc{Kimata}$^2$,  Michal \textsc{Vali\v{s}ka}$^{1}$, Fuminori \textsc{Honda}$^3$, Dexin \textsc{Li}$^3$, Daniel \textsc{Braithwaite}$^{1}$, G\'{e}rard \textsc{Lapertot}$^{1}$, William \textsc{Knafo}$^4$, Alexandre \textsc{Pourret}$^{1}$, Yoshiki J. \textsc{Sato}$^3$, Yusei \textsc{Shimizu}$^3$, Takumi \textsc{Kihara}$^2$, Jean-Pascal \textsc{Brison}$^{1}$, Jacques \textsc{Flouquet}$^{1}$, Dai \textsc{Aoki}$^{1,3}$ }
\inst{$^1$Univ. Grenoble Alpes, CEA, IRIG-Pheliqs, 38000 Grenoble, France  \\
$^2$Institute for Materials Research, Tohoku University, Sendai 980-8578, Japan\\
$^3$Institute for Materials Research, Tohoku University, Ibaraki 311-1313, Japan\\
$^4$Laboratoire National des Champs Magn\'{e}tiques Intenses, UPR 3228, CNRS-UPS-INSA-UGA, 143 Avenue de Rangueil, 31400 Toulouse, France\\
}

\abst{We studied the anisotropy of the superconducting upper critical field $H_{\rm c2}$ in the heavy-fermion superconductor UTe$_2$ under hydrostatic pressure by magnetoresistivity measurements. In agreement with previous experiments we confirm that superconductivity disappears near a critical pressure $p_{\rm c} \approx 1.5$~GPa, and a magnetically ordered state appears. The unusual $H_{\rm c2}(T)$ at low temperatures for $H \parallel a$ suggests that the multiple superconducting phases which appear under pressure have quite different $H_{\rm c2}$. For a field applied along the hard magnetization $b$ axis $H_{\rm c2} (0)$ is glued to the metamagnetic transition $H_{\rm m}$ which is suppressed near $p_{\rm c}$. The suppression of $H_{\rm m}$ with pressure  follows the decrease of temperature $T_{\chi}^{\rm max}$, at the maximum in the susceptibility along $b$. The strong reinforcement of $H_{\rm c2}$ at ambient pressure for $H \parallel b$ above 16~T is rapidly suppressed under pressure due to the increase of $T_{\rm sc}$ and the decrease of $H_{\rm m}$. The change in the hierarchy of the anisotropy of $H_{\rm c2}(0)$ on approaching $p_{\rm c}$ points out that the $c$ axis becomes the hard magnetization axis. }

\begin{document}
\maketitle

In many strongly correlated electron systems unconventional superconductivity (SC) appears close to a quantum phase transition where a long range ordered phase is suppressed by tuning a control parameter of the system such as pressure, doping, or charge carrier number.\cite{Flouquet2006,Lohneysen2007}    It is believed that the enhancement of the magnetic and electronic fluctuations are the glue for the superconducting pairing. This has been most impressively shown for ferromagnetic superconductors such as URhGe and UCoGe, where the pairing strength itself can be tuned by the magnetic field \cite{Miyake2008, Wu2017, Aoki2019a}. A magnetic field applied along the easy magnetization axis of these orthorhombic systems lowers the superconducting pairing, while a magnetic field applied along the hard magnetization axis enhances the superconducting pairing strength as the field drives the system to a collapse of the ferromagnetism. This enhancement of the pairing strength for $H \parallel b$ results in the reentrance of SC in the field range of 9~T - 12~T  in URhGe \cite{Levy2005} and an enhancement of SC in UCoGe.\cite{Aoki2009}  
The strong internal exchange field in these ferromagnetic superconductors and the extremely high ratio of the upper critical field $H_{c2}$  compared to the superconducting transition temperature  $T_{\rm sc}$ imposes a non-unitary spin triplet state with equal spin pairing (ESP) which is a superconducting state very rarely realized in bulk materials. 

Recently the superconducting state below $T_{\rm sc} =1.6$~K of the paramagnetic heavy fermion compound UTe$_2$ has also been proposed to be a spin triplet
superconductor.\cite{Ran2019, Aoki2019} Evidence for this is obtained from the very large and strongly anisotropic $H_{\rm c2}$,  which is 6~T for the field applied along the easy magnetization axis $a$, 11~T for the $c$-axis but is extremely field-enhanced up to 35~T for field along the hard magnetization $b$-axis (exceeding by far the Pauli limitation for a singlet superconductor). Above 35~T, where a metamagnetic transition with a huge jump of the magnetization  $M$ occurs,\cite{Miyake2019, Ran2019a} 
SC is abruptly suppressed. 
An additional singularity of UTe$_2$ is that under pressure multiple superconducting phases occur.\cite{Braithwaite2019}   $T_{\rm sc}$ is initially linearly suppressed with pressure, but for $p > 0.3$~GPa two specific heat anomalies are observed and the upper superconducting anomaly  increases up to 3~K while the lower continues decreasing in temperature. In that experiment SC is suppressed near 1.7~GPa and a new magnetically ordered phase appears. The increase of $T_{\rm sc}$ by a factor of 2 compared to the the ambient pressure value and the collapse of the superconducting regime near 1.7~GPa has been confirmed by resistivity experiments.\cite{Braithwaite2019, Ran2019b, Lin2020}

In the present work we concentrate on the anisotropy of $H_{\rm c2}$ under pressure. First we present the pressure dependence of the susceptibility.  We show that (i) $H_{\rm c2}$  for $H \parallel a$ is unusally  enhanced at low temperature, (ii) for $H \parallel b$ the superconducting phase is suppressed at a metamagnetic transition, which  decreases in field by increasing pressure, and (iii) the upper critical field for $H \parallel c$ crosses that for $H \parallel b$ for $p > 1.2$~GPa. 
This could be related to a change in the magneto-cristalline anisotropy and we speculate that the $c$ axis becomes the hardest magnetization axis at high pressure. 

Single crystals of UTe$_2$ have been grown by the chemical vapor transport method as described previously, in Grenoble and in Oarai.\cite{Aoki2019, Aoki2020_sces} High pressure susceptibility measurements have been performed in a Quantum design MPMS using a specially designed Cu-Be piston cylinder cell. For the resistivity experiments under pressure in Grenoble, a several millimeter long needle-like single crystal has been cut in three pieces and mounted in a piston cylinder cell. So  field could be applied on different parts of the same crystal along the $a$, $b$, and $c$ axes. The current was injected along the field direction for $H\parallel a$ while for the other directions the current is perpendicular to the field. 
The measurements have been performed in a Quantum design PPMS (maximal field 9~T) and a dilution refrigerator ($H_{\rm max} = 13$~T). In parallel, high pressure experiments have been performed in Oarai using an Oxford top-loading dilution refrigerator ($H_{\rm max} = 15$~T). Results from experiments in Grenoble and Oarai are very similar, except that the samples used in Oarai have a slightly higher $T_{\rm sc}$ under pressure. In addition we performed, at selected pressures, measurements up to 30~T in the high field laboratory in Sendai using $^3$He and $^4$He cryostats.  

\begin{figure}[t]
\includegraphics[width=1\columnwidth]{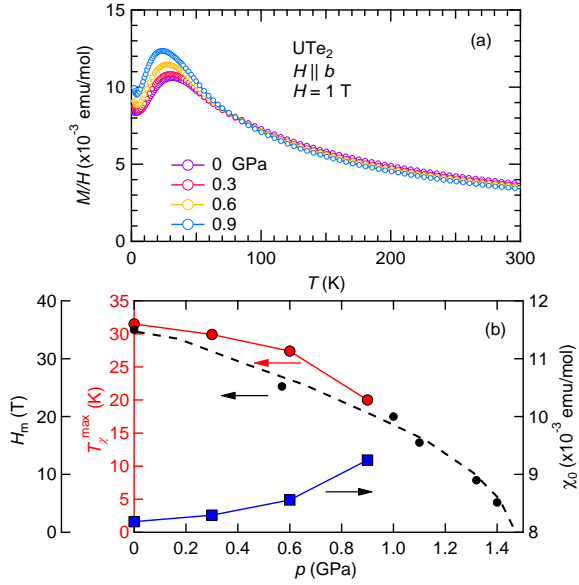}
\caption{(Color online) (a) Temperature dependence of $M/H$ for $H=1$~T applied along the $b$ axis of UTe$_2$ at different pressures. The small upturn below 4~K is due to the background contribution from the pressure cell. (b) Pressure dependence of the temperature of the maximum of the susceptibility $T_{\chi}^{\rm max}$ (left red  scale, red circles), $M/H$ extrapolated to $T=0$ (blue squares, right scale) and $H_{\rm m}$ (left scale, black circles). }
\label{Fig1}
\end{figure}

In Fig.~\ref{Fig1}(a) we show the magnetic susceptibility as $M/H$  for a field of 1~T applied along the $b$ axis as a function of temperature for different pressures up to 0.9~GPa. At zero pressure, the susceptibility has  a maximum at $T_{\chi}^{\rm max} \approx 31.5$~K, slightly lower than outside the pressure cell. 
 At zero pressure  $T_{\chi}^{\rm max}$ is linked to the metamagnetic transition at $H_m \approx 35$~T. \cite{Miyake2019, Knafo2019} It shows that the same energy scale is responsible for the formation of  a correlated electronic regime in zero magnetic field  and pushes the system under magnetic field to a metamagnetic transition\cite{Aoki2013}. In UTe$_2$ a huge jump of  the magnetization $\Delta M = 0.6~\mu_{\rm B}$ is observed at $H_{\rm m}$ at $p=0$, while the susceptibility $\partial M/\partial H$ is almost unchanged below and above $H_{\rm m}$. The maximum of the susceptibility $T_{\chi}^{\rm max}$  decreases under pressure, and at 0.9~GPa we find $T_{\chi}^{\rm max} \approx $ 20~K. The absolute value of $M/H$ at low temperature is inversely proportional to $T_{\chi}^{\rm max}$ and increases under pressure [see Fig.~\ref{Fig1}(b)]. We have also added the pressure dependence of the metamagnetic transition field $H_{\rm m}$ detected by magnetoresistivity. Importantly it follows the pressure dependence of  $T_{\chi}^{\rm max}$. A rough extrapolation yields $H_{\rm m} \to 0$ near 1.5~GPa, i.e.~near the pressure where SC is replaced by a magnetically ordered state at $p_c$.\cite{Braithwaite2019, Ran2019b, Lin2020}

\begin{figure}[t]
\begin{center}
\includegraphics[width=0.8\columnwidth]{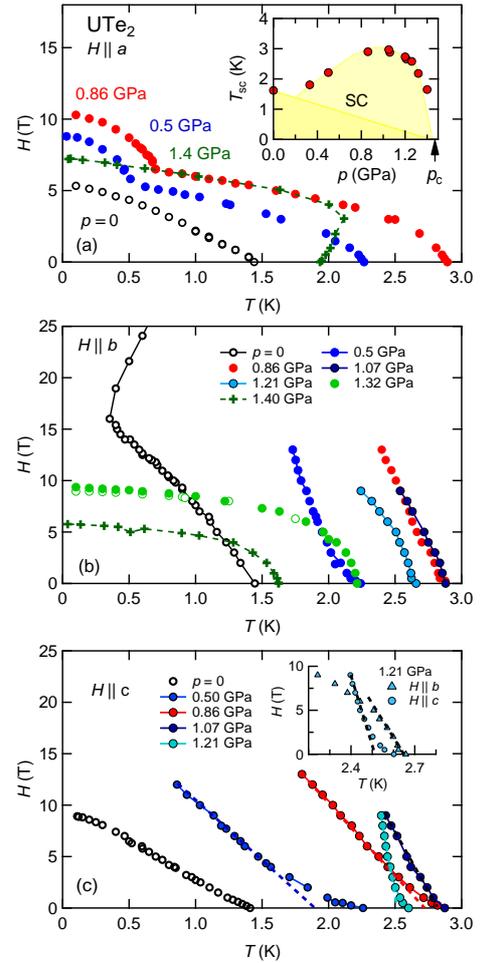}
\caption{(Color online) Upper critical field $H_{\rm c2}$ for a magnetic field applied along different directions: (a) $H \parallel a$, (b) $H\parallel b$ and (c) $H\parallel c$ axis. Open circles are taken from Ref.~\citen{Knebel2019}. Full circles are from measurements in Grenoble, crosses give $H_{\rm c2}$ measured at 1.4 GPa in Oarai.  The difference in $T_{\rm sc}$ at 1.4~GPa at zero field is probably due to the slight pressure inhomogeneity in the cell. For $H\parallel a$ we find a strong enhancement of $H_{\rm c2}$ at low temperatures indicating the possibility of multi-superconducting phases. The insert in (a) shows $T_{\rm sc}$ as a function of pressure. Dashed lines in (c) are an extrapolation of $H_{\rm c2}$ from high fields. The insert in (c) shows the crossing of  $H_{\rm c2}$ for $H \parallel b$ and $\parallel c$ at 1.2 GPa.   We used the criterion $\rho = 0$  to determine $H_{\rm c2}$ in all cases.}
\label{Fig2}
\end{center}
\end{figure}

Figure \ref{Fig2} displays $H_{\rm c2}$ as a function of temperature along the $a$, $b$ and $c$ axes of UTe$_2$. $T_{\rm sc}(H)$ or $H_{\rm c2}(T)$ have been determined from temperature sweeps at constant field, or field sweeps at constant temperature using the criterion $\rho =0$ (see Supplemental Material for raw data)\cite{suppl}. The inset in Fig.~\ref{Fig2}(a) shows the pressure-temperature phase diagram. We find good agreement with the previous reports.\cite{Braithwaite2019, Ran2019b} No complete superconducting transition has been found at 1.48 GPa. Thus, compared to Ref.\citen{Braithwaite2019}, the critical pressure $p_{\rm c}$ up to which SC can be observed is slightly lower. We also do not observe the initial decrease of $T_{\rm sc}$,\cite{Braithwaite2019} as our first pressure is already 0.3 GPa. 
%

$H_{\rm c2}(T)$ for $H \parallel a$ is shown in Fig.~\ref{Fig2}(a).  The ambient pressure data are on a different sample. At zero temperature $H_{\rm c2}(0) \approx 6$~T. At the same field a Lifshitz transition of the Fermi surface is observed \cite{Niu2020}. Increasing the pressure, $T_{\rm sc}$ increases with a maximum at $p \approx 1$~GPa. Most spectacularly we find a strong enhancement of $H_{\rm c2}$ below 0.6~K at 0.5~GPa and below 0.75~K at 0.86~GPa marked by a kink in $H_{\rm c2}(T)$. 
In some organic superconductors or iron-pnictides a FFLO state is observed at high field \cite{Wosnitza2019, Cho2017, Kasahara2020}. Here this is certainly not the case, as a FFLO state induces at most a positive curvature of $H_{\rm c2}$, not such a kink, leading to a maximum increase of $H_{\rm c2}(0)$ of around $6\%$ for the pure paramagnetic limit of three dimensional superconductors, not $25\%$ as observed here.
It is most likely due to multiple  superconducting phases with different order parameters,  as in the phase diagrams of  UPt$_3$, or Th-doped UBe$_{13}$ \cite{Hasselbach1989, Fisher1989, Shimizu2017}.Thermodynamic measurements are required to reveal the possible appearance of extra superconducting phases below the superconducting boundary detected here \cite{Aoki2020}.

The strong curvature of  $H_{\rm c2} (T)$ under pressure at 0.5 GPa and 0.86 GPa points to the presence of a strong Pauli paramagnetic effect. Even at $p=0$, $H_{\rm c2}(T)$ determined specific heat measurements points to a paramagnetic limitation.\cite{Kittaka2020}
At 1.4~GPa $H_{\rm c2}$ has clearly a reentrant behavior at low field. Increasing the magnetic field, $H_{\rm c2}$ increases from 1.93~K at $H = 0$ to 2.1~K at $3$~T. 
It shows that close to the critical pressure, field is enhancing (or restoring) SC in this direction. At lower temperature, we do not see any enhancement of SC at 1.4~GPa in difference to the lower pressures.  

In Fig.~\ref{Fig2}(b) we show $H_{\rm c2}(T)$ for $H \parallel b$.  The known remarkable feature at $p=0$ is the enhancement of SC from 17~T up to 35~T \cite{Knebel2019}. In the Grenoble experiment, the maximum field has been 13~T, such that we could not follow the field enhancement. However, we clearly see the change of behavior of $H_{\rm c2}(T)$ as a function of pressure near the maximum of $T_{\rm sc}$ at $p\approx 1$~GPa. While at 0.86GPa and 1.07~GPa $H_{\rm c2}(T)$ is almost linear up to the highest field, we observe, for $p  \gtrapprox 1.21$~GPa, a marked downward curvature of $H_{\rm c2}$ near $T_{sc}$. 
This strong curvature of $H_{\rm c2} (T)$ observed on approaching $p_c$, suggests either a strong Pauli paramagnetic limit, or a strong field dependence of the pairing.  As we will see below another underlying effect is  the suppression of the metamagnetic transition. Moreover, at 1.32~GPa,
we can detect a clear hysteresis of 0.5~T between field sweep up and field sweep down, for temperatures below 1.2~K (roughly 0.5$T_{\rm sc}$ ) [closed and open circles in Fig.~\ref{Fig2}(b) and Supplemental Material], indicating a first order nature of the superconducting transition.

In Fig.~\ref{Fig2}(c) we plot $H_{\rm c2}(T)$ for $H\parallel c$ for different pressures. Strikingly, we see upward curvatures very close to $T_c$ for all pressures. It is most pronounced at 0.5~GPa. Such an initial upward curvature may occur in resistivity experiments due to filamentary SC or 
to multiband effects, with dominant pairing for light carriers. Similarly, in Ref.~\citen{Braithwaite2019} or in UCoGe\cite{Wu2018} such behaviour is reported from resistivity measurements. On the contrary, $H_{\rm c2}$ determined by ac calorimetry for $H \parallel c$ has a linear temperature dependence at $T_{\rm sc}$\cite{Braithwaite2019}. Here, we observe that for fields above 2~T, $H_{\rm c2}(T)$ varies linearly with temperature and the slope is strongly enhanced with pressure. In  the inset of  Fig.~\ref{Fig2}(c) we compare $H_{\rm c2}$ for the $b$ axis and $c$ axis at 1.21~GPa. While for the $b$ axis $H_{c2}$ has the usual downward curvature near $T_{\rm sc} (0)$, it has an upward curvature for the $c$ axis and is linear only above 5~T. Neglecting the initial curvature,  the slope  for $H \parallel c$ is about $-90$ T/K at 1.21 GPa, comparable to that determined from  ac calorimetry close to the critical pressure\cite{Braithwaite2019}. This indicates an extreme re-inforcement of SC along the $c$ axis close to $p_{\rm c}$\cite{noteRan}.

\begin{figure}[t]
\begin{center}
\includegraphics[width=.8\columnwidth]{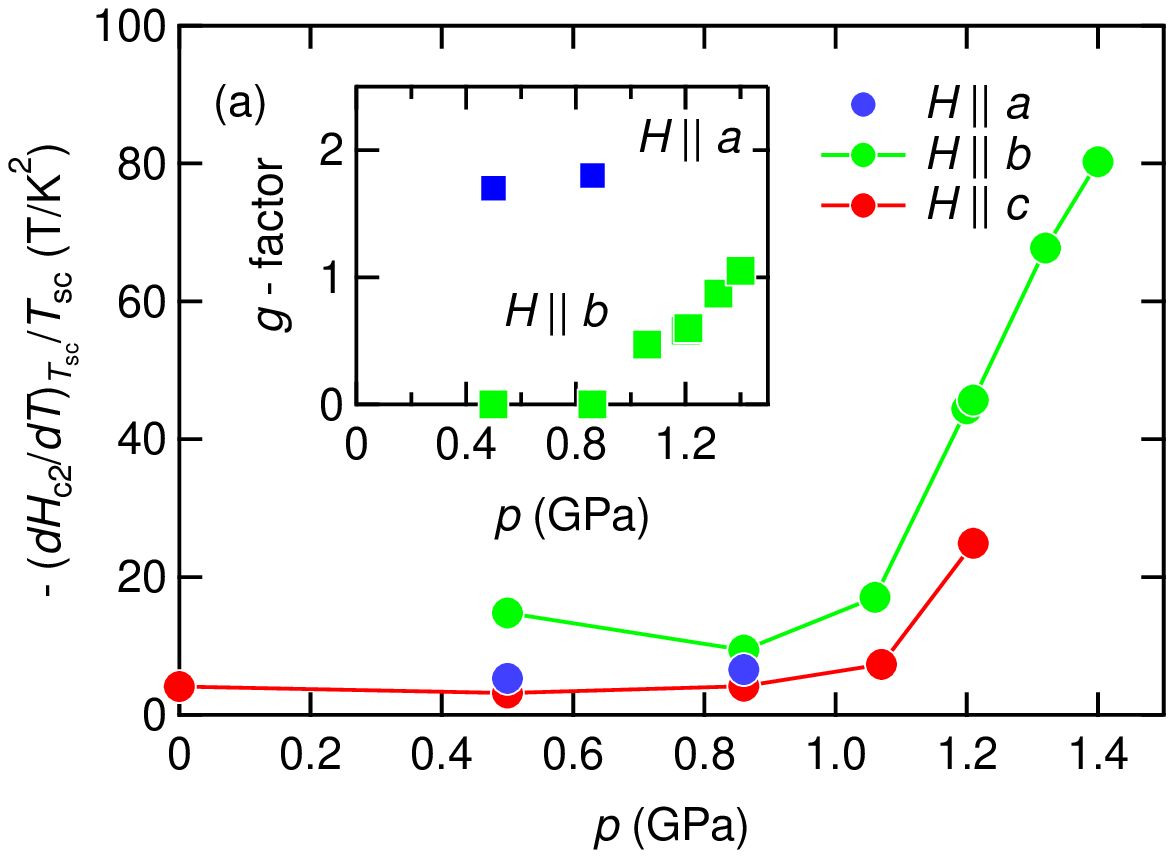}
\includegraphics[width=.75\columnwidth]{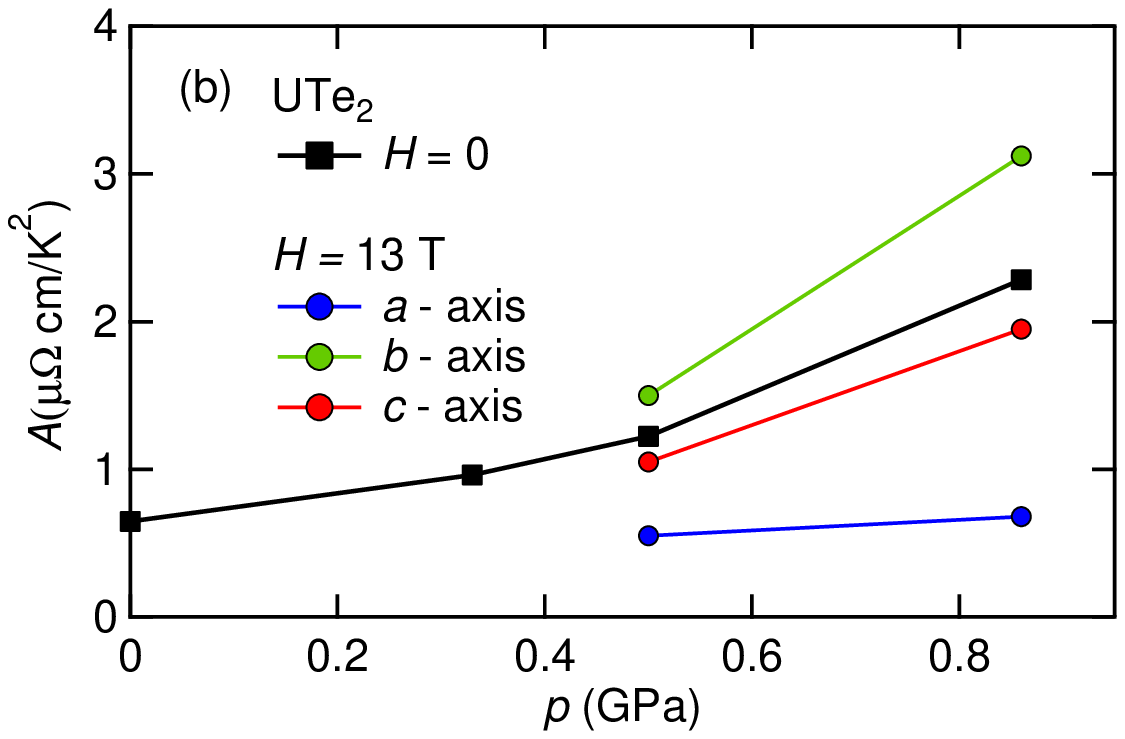}
\caption{(Color online) (a) Pressure dependence of the initial slope of $H_{\rm c2}$ at $T_{\rm sc}$ normalized by $T_{\rm sc}$ for the  three crystallographic axes. Inset: Pressure dependence of the $g$ factor indicating the importance of the paramagentic limitation of $H_{\rm c2}$. (b) Pressure dependence of the $A$ coefficient of the resistivity at $H =0$ (squares). We have also added the variation of the $A$ coefficent at 13~T for $H \parallel a$, $b$, and $c$ axes.   }
\label{Fig3}
\end{center}
\end{figure}

Closer insights on the superconducting properties can be obtained by analysing in more details  $H_{\rm c2}(T)$ close to $T_{\rm sc}$. 
In order to extract as precisely as possible the initial slope $H_{\rm c2}'  = (dH_{\rm c2}/dT)_{T=T_{\rm sc}} $ even, when there is a strong positive curvature, we made a (weak-coupling) fit of the data taking into account both the orbital  and a possible Pauli limitation [see Fig. ~S9 in Supplemental Material]. In Fig.~\ref{Fig3}(a) we plot this determination of the initial slope $H_{\rm c2}'$ normalised by  $T_{sc}$ as a function of pressure for the three crystal directions. 
In a normal superconductor, i.e.~when the pairing strength does not change significantly with the applied field, it is a good measure of the average Fermi velocity $v_{\rm F}$ in directions perpendicular to the applied field, as  $H_{\rm c2}' \propto -T_{\rm sc} / v_{\rm F}^2$. So, it is roughly proportional to the square of the corresponding effective mass $(m^\star)^2$.  
In Fig.~\ref{Fig3}(a) we plot $-H_{\rm c2}' / T_{\rm sc}$ as function of pressure. While up to 1~GPa the  $H_{\rm c2}'/T_{\rm sc}$  is almost constant, it increases strongly above 1~GPa. For the $b$ axis we find an increase by a factor 9 between 1~GPa and 1.4~GPa. From Ref.~\citen{Braithwaite2019} we know that $H_{\rm c2}'$ increases for the $c$ axis until SC is suppressed. Our data here are in good agreement with those determined by ac-calorimetry, and the increase up to $p_{\rm c}$ for $H \parallel c$ is by a factor 6.
For $H\parallel a$ we plot the data only up to 0.86~GPa, but at 1.4~GPa the initial slope has already changed sign. Thus, $H_{\rm c2}'$ is strongly reinforced on approaching $p_{\rm c}$ for all three directions. 
 
Of course, the initial slope is not a direct measure of the effective mass $m^\star$ due to the possible field enhancement of the strong-coupling parameter $\lambda$ \cite{Wu2017,Ran2019,Knebel2019}. On the other hand, such a large increase of $H_{\rm c2}'$ for $H \parallel b$ certainly gives the right trend for the pressure increase of $m^\star$. 
The pressure dependence of $H_{\rm c2}'$ compares qualitatively with the pressure dependence of the $A$ coefficient of the $T^2$ term of the resistivity. At low pressure the resistivity follows well a $\rho(T) = \rho_0 +AT^2$ Fermi-liquid temperature dependence, and with increasing pressure the coefficient increases from $A = 2.7  \mu\Omega$~cm/K$^2$ to 17.5~$\mu\Omega$~cm/K$^2$ at 0.86~GPa by a factor 5 at zero field, see Fig.~\ref{Fig3}(b). We have also plotted the $A$ coefficient for a field of 13~T which shows that for the $b$ and $c$ axes it is already enhanced, while for the $a$ axis $A$ is almost constant.  However, above  1~GPa the resisitivity is almost linear in the normal state above the superconducting transition indicating the closeness to some quantum critical point (see also Supplemental Material). 

In the inset of Fig.~\ref{Fig3}(a) we have plotted the electronic $g$ factor determined from fitting also the curvature of $H_{\rm c2} (T)$, which is related to the Pauli limitation $H_{\rm c2}^{\rm P}(0) = \frac{\sqrt{2}\Delta}{g\mu_B}$\cite{Clogston1962}. Of course, these values are only indicative, because neither strong coupling effects nor a field dependence of the pairing have been taken into account. However, it is clear that for $H \parallel a$ and above 1~GPa for $H \parallel b$, the Pauli limitation of $H_{\rm c2}$ is not negligible. The fact that paramagnetic limitation appears along two perpendicular directions at least puts strong constraints on the possible superconducting order parameters: for example, it is not possible to have a simple real $d$-vector fixed in a given direction as it would lead to a paramagnetic limitation only along this direction (and ESP states in the perpendicular directions). It could be that we have a complex $d$-vector (non-unitary state)\cite{Metz2019,Kittaka2020} with no component along the $c$ axis, or an $A_{1u}$ order parameter (like in the $B$ phase of superfluid $^3$He), with a paramagnetic limitation present along the three directions.



\begin{figure}[t]
\includegraphics[width=0.9\columnwidth]{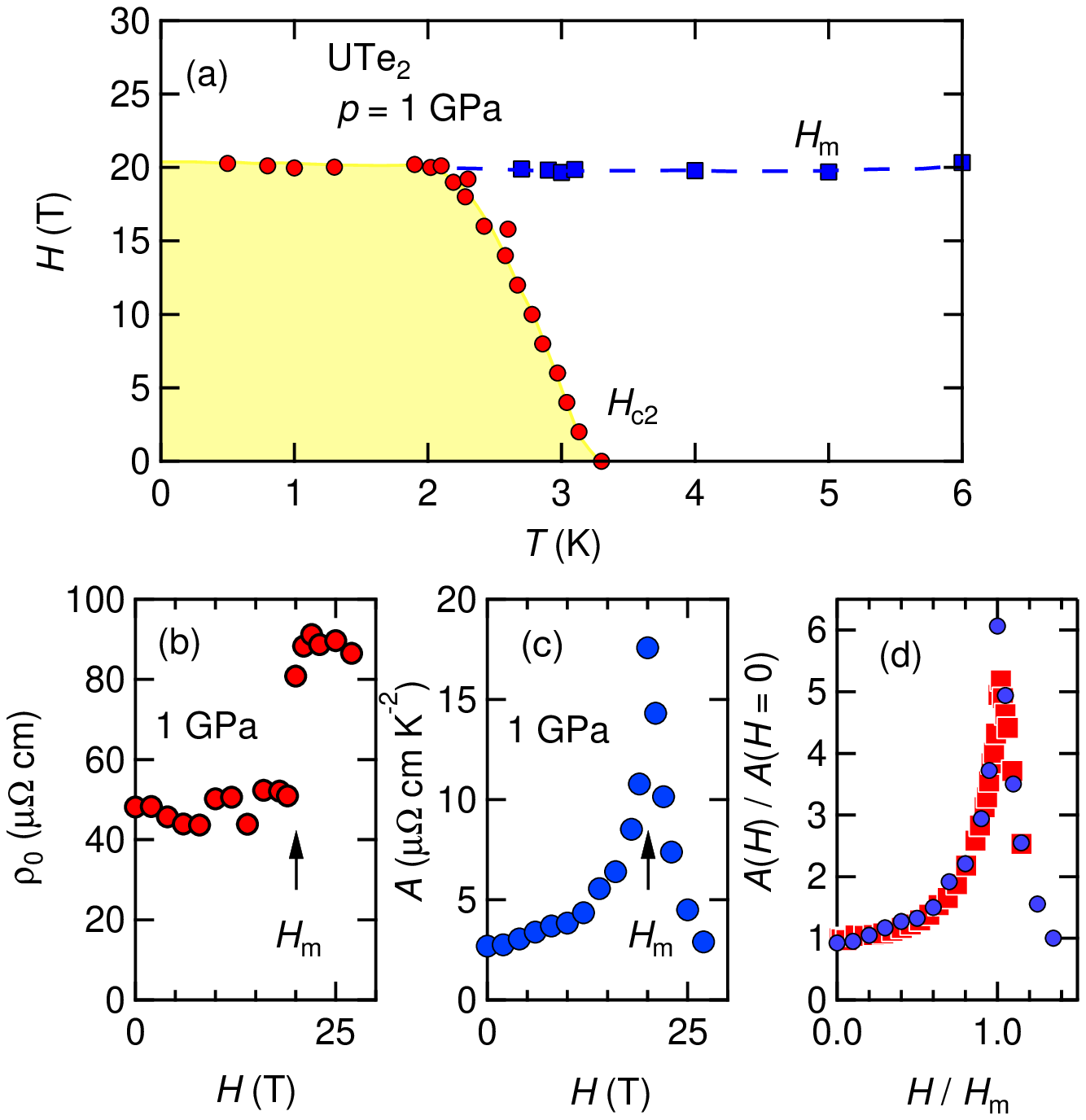}
\includegraphics[width=0.9\columnwidth]{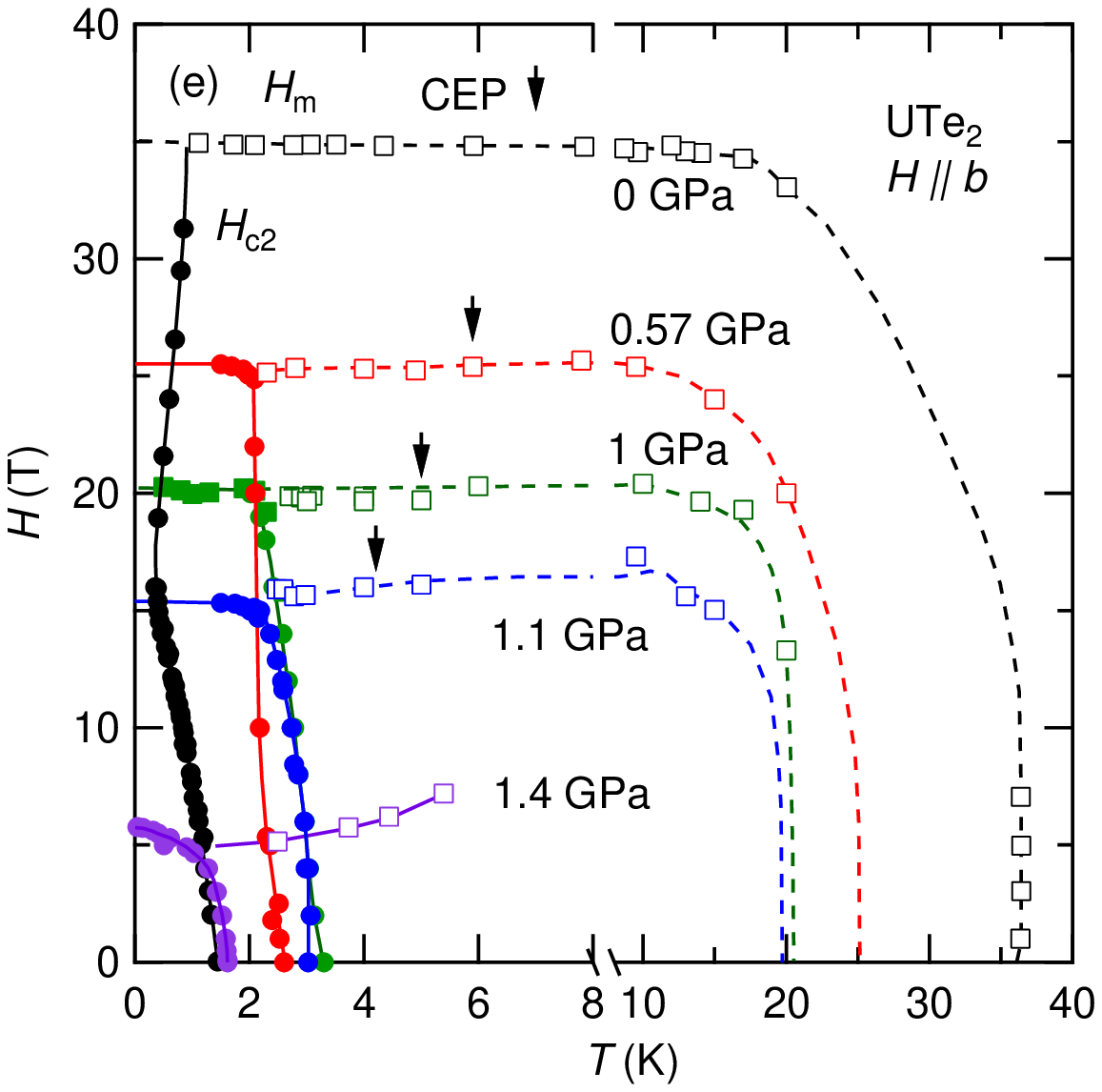}
\caption{(Color online) (a) Magnetic and superconducting phase diagram for $p=1$~GPa for $H \parallel b$. SC is abruptly suppressed for $H > H_{\rm m}$. (b) Field dependence of the residual resistivity $\rho_0 (H)$, which increases suddenly by a factor 2 at $H_{\rm m}$. (c) Field dependence of the $A$ coefficient, which is strongly enhanced at $H_{\rm m}$. (d) Comparison of the $A$ coefficient at 1~GPa (blue) to that at $p=0$ (red squares). (e) $H,T$ phase diagrams ($H\parallel b$) for different pressures, full symbols correspond to $H_{\rm c2}$, open squares to $H_{\rm m}$ or a crossover above the critical end point (CEP), which is indicated by the arrows. Dashed and solid lines are guide to the eye. 
 }
\label{Fig4}
\end{figure}

Finally, in Fig.~\ref{Fig4}(a) we show the phase diagram for $H\parallel b$ at 1~GPa. We clearly see that SC is suppressed continuously with field, but is suddenly cut off when $H_{\rm c2}$ is of the same order as the metamagnetic transition field $H_{\rm m}$.  At 1~GPa we find $H_{\rm m} \approx 20$~T, in very good agreement with the maximum in the temperature dependence of the susceptibility ($T_{\chi}^{\rm max}   \approx 20$~K) [see Fig.~\ref{Fig1}(b)]. The first order nature of the transition at $H_{\rm m}$ is clearly manifested in the jump of the residual resistivity $\rho_0$ at $H_{\rm m}$ shown in Fig.~\ref{Fig4}(b) from about $50~\mu\Omega$cm to 90~$\mu\Omega$cm. The jump  $\Delta \rho_0 \approx 40 \mu\Omega$cm is three times smaller than at $p=0$. However, there are also strong magnetic fluctuations developing at $H_{\rm m}$ connected to a huge increase of the $A$ coefficient. As shown in Fig.~\ref{Fig4}(c) the relative field dependence of $A(H)$ at 1~GPa scales perfectly with the $A(H)$ at $p=0$ [see Fig.~\ref{Fig4}(d)].   The disappearance of SC right above $H_{\rm m}$ is in contrast to the rather symmetric field dependence of $A (H)$.
In Fig.~\ref{Fig4}(e) we have summarized the $H,T$ phase diagrams for $H \parallel b$ for different pressures. We want to stress that the reinforcement of $H_{\rm c2}$ observed at zero pressure above 16~T occurs near 8~T at $p=0.57$~GPa  and has completely disappeared at 1~GPa. If it is due to a change in the superconducting order parameter,\cite{Ishizuka2019} only the high field phase will survive close to $p_c$.\cite{Lin2020} The signature of the metamagnetic transition gradually fades out between 1~GPa and 1.4~GPa, where only a kink in the resistivity is observed, that could be  a signature of $H_{\rm m}$. This explains why no clear cut off of SC at $H_{\rm m}$ appears close to $p_c$. 

Metamagnetism in heavy fermion compounds leads generally to a drastic change of the nature of the magnetic correlations with a reconstruction of the Fermi surface. For example in ferromagnetic superconductors this is well established in URhGe for field applied along the $b$ axis\cite{Gourgout2016}. Similar phenomena have been detected on crossing the antiferromagnetic to paramagnetic transition of CeRh$_2$Si$_2$ \cite{PalacioMorales2015} or in UPd$_2$Al$_3$ \cite{Gourgout_these}. A highly studied case is the CeRu$_2$Si$_2$ series (see e.g.\cite{Flouquet2006, HAoki2014});  again, for pure CeRu$_2$Si$_2$,  the sharp crossover at $H_{\rm m}$ from a nearly antiferromagnetic phase to a polarized paramagnetic phase, is associated with a drop of antiferromagnetic correlations above $H_{\rm m}$ and a drastic Fermi surface reconstruction. A simple "rule of thumb" is that ferromagnetic or antiferromagnetic correlations drop above $H_{\rm m}$ and mainly only local fluctuations survive.

Thus in UTe$_2$ if pairing is indeed due to the FM intersite correlations, it should be strongly suppressed above $H_{\rm m}$ : it is clear for $H \parallel b$ that it is strongly suppressed. Furthermore even at $H = 0$ above $p_{\rm c} \approx 1.5$~GPa, it is still not known whether UTe$_2$ shows ferromagnetic or antiferromagnetic order (there is no clear  indication for ferromagnetism in the transport measurements)
The anisotropy of $H_{\rm c2}$  close to $p_{\rm c}$ is reversed compared  to that at $p = 0$, and we find $H_{\rm c2}^{c}(0) > H_{\rm c2}^{a}(0)\sim H_{\rm c2}^{b}(0)$ for the $c$, $a$, and $b$ axes, respectively. As the anisotropy of the $H_{\rm c2}(0)$ in most heavy fermion  superconductors (UPt$_3$ might be the only exception) follows that of the susceptibility, it is likely that  the $c$ axis becomes the hard axis near the critical pressure.

Our study of $H_{\rm c2}$ under pressure for the field along the $a$, $b$, and $c$ axes  reveals unique features of the superconducting phase of UTe$_2$, which clearly demands further thermodynamic experiments. Magnetic field and pressure induced  changes of the ferromagnetic interactions as well as possible  drastic Fermi surface reconstructions at $H_m$ are important ingredients. The duality between the local and itinerant character of the 5$f$ electrons is important to understand the change in the magnetocrystalline energy leading to a suppression of the metamagnetic field under pressure. SC also evolves strongly in UTe$_2$. For $H\parallel a$ we have shown that $H_{\rm c2}$ is unusually enhanced at low temperatures suggesting multiple superconducting phases.\cite{Aoki2020} 
The upper critical field for the $b$ axis is determined by the mutual balance of the increase of $T_{\rm sc}$ and the decrease of $H_{\rm m}$ which cuts off SC  at the metamagnetic critical field $H_{\rm m}$. Close to the collapse of SC near 1.5~GPa $H_{\rm c2}(0)$ is highest along the $c$ axis.

\begin{acknowledgment}
We thank K.~Machida, Y.~Yanase, V.~P.~Mineev and Z.~Zhitomirsky for fruitful discussions. 
We acknowledge the financial support of the Cross-Disciplinary Program on Instrumentation and Detection of CEA, the French Alternative Energies and Atomic Energy Commission,  KAKENHI (JP15H05882, JP15H05884, JP15K21732, JP16H04006, JP15H05745, JP19H00646, JP19K03736) and GIMRT (19H0416, 19H0414). \\
During finishing the present draft, quite similar results for $H \parallel b$ to those described here have been put on arXiv (see Ref.~\citen{Lin2020}).
\end{acknowledgment}


\newpage

{\bf Supplemental Material}
\vspace{0.5cm}

In this Supplemental Material we show additional data to those shown in the main article, further data will be made available on special demands. 

\section{Resistivity under high pressure}

\begin{figure}[b]
\makeatletter
\renewcommand{\thefigure}{S\@arabic\c@figure}
\makeatother
\begin{center}
\includegraphics[width=0.8\columnwidth]{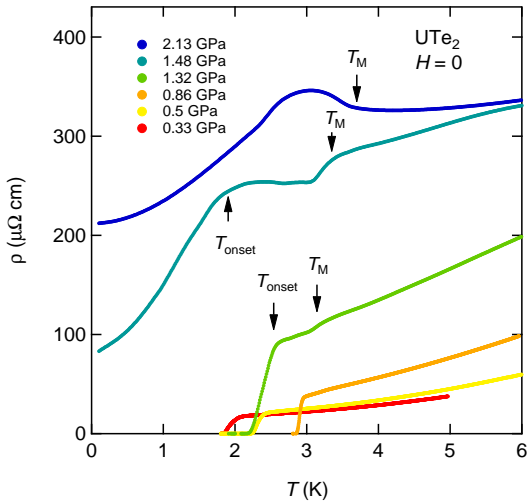}
\caption{Resistivity as a function of temperature of UTe$_2$ at zero magnetic field for various pressures. }
\label{sup_rho}
\end{center}
\end{figure}

Figure~\ref{sup_rho} shows the temperature dependence of the  resistivity of UT$_2$ at zero field for different pressures. Up to a pressure of 1.32~GPa zero resisitivty is observed below the superconducting transition temperature $T_{\rm sc}$. For higher pressures the resistivity at lowest temperature is finite, i.e.~no complete superconducting transition is observed. $T_{\rm onset}$ indicates the onset of the superconducting transition. At 1.48 GP the transition is not complete and extremely broad. This indicates that the slope of $dT_{\rm sc}/dp$ is very steep. In addition, already at 1.32~GPa we observe a kink in the resistivity above the onset of the superconducting transition. This is even more pronounced at 1.48~GPa and may be linked to the magnetically ordered state which is observed for higher pressures. This may indicate that the transition from the a superconducting ground state to the magnetically ordered state is first order. We want to emphesize that at 2.13 GPa, the magnetic transition is not at all like expected for a simple ferromagnetic or antiferromagnetic transition and deserves detailed investigatios in future. The shape of the resistivity curves is different than those published in Ref.~\citen{Ran2019b}, mainly due to the different direction of the current, which is along the $a$ axis in the present experiment. 

\newpage
\section{Resistivity at $p = 0.86$~GPa}

\begin{figure}[t]
\makeatletter
\renewcommand{\thefigure}{S\@arabic\c@figure}
\makeatother
\begin{center}
\includegraphics[width=0.8\columnwidth]{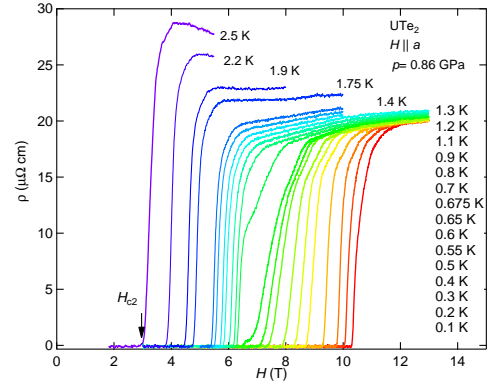}
\caption{Magnetoresistance for magnetic field applied along the $a$ direction at $p = 0.86$~GPa. The arrow indicate the criterion $\rho =0$ used to determine $H_{\rm c2}$. }
\label{sup_fig1}
\end{center}
\end{figure}

\begin{figure}[h]
\makeatletter
\renewcommand{\thefigure}{S\@arabic\c@figure}
\makeatother
\begin{center}
\includegraphics[width=0.8\columnwidth]{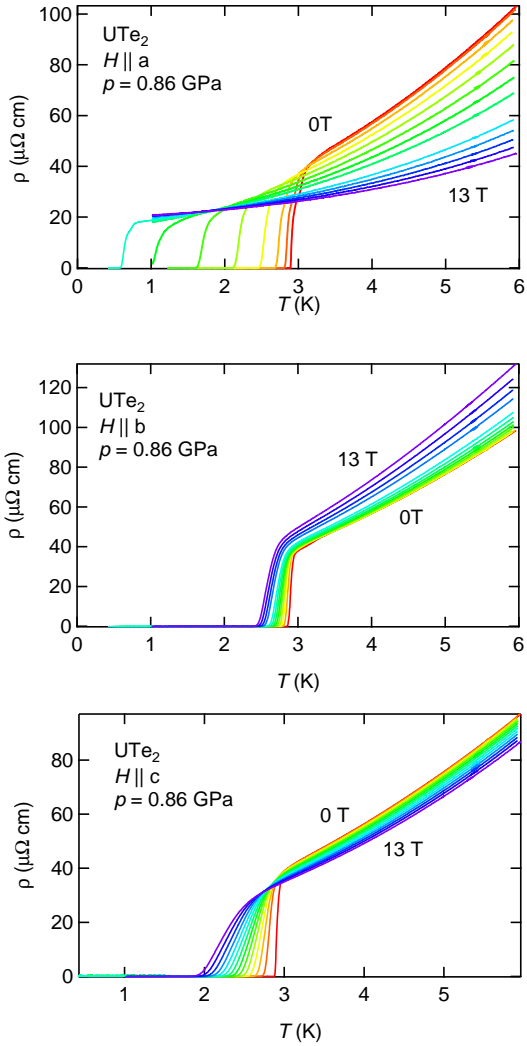}
\caption{Temperature dependence of the resistivity for fields between 0~T and 13~T at $p = 0.86$~GPa for the $H \parallel a$ (upper panel),   $H \parallel b$ (middle panel), $H \parallel c$ (lower panel). }
\label{sup_fig2}
\end{center}
\end{figure}

\begin{figure}[h]
\makeatletter
\renewcommand{\thefigure}{S\@arabic\c@figure}
\makeatother
\begin{center}
\includegraphics[width=0.8\columnwidth]{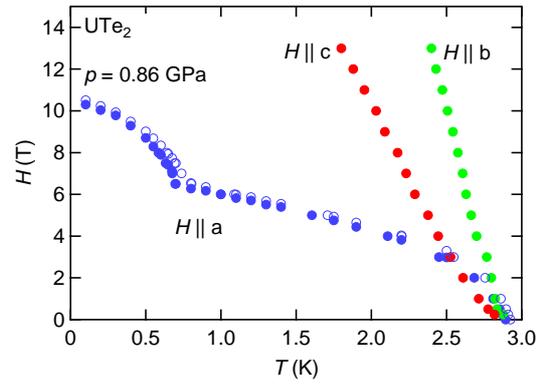}
\caption{Upper critical field $H_{\rm c2}$ at $p=0.86$~GPa for the three principal crystallographic directions. Open circles corresponds to the mid-point of the transition for $H \parallel a$. }
\label{sup_fig3}
\end{center}
\end{figure}

Figure~\ref{sup_fig1} shows the magnetoresistance $\rho (H)$ as a function of field applied along the $a$ axis of the orthorhombic UTe$_2$ at $p=0.86$~GPa for field along the $a$ axis.  The upper critical field $H_{\rm c2}$ has been defined by the $\rho = 0$ criterion.  In Fig.~\ref{sup_fig2} we show the temperature dependence of the resistivity for the three crystallographic directions for fields between 0~T and 13~T, the current is always injected along the $a$ axis. We checked carefully that there is no dependence on the criterion used to determine $H_{\rm c2}$ on the overall shape of the upper critical field for $H \parallel a$ as shown in Fig.~\ref{sup_fig3}. Most remarkably is the strong enhancement of $H_{\rm c2}$ below 700~mK for $H\parallel a$. These strong increase has also been verified on different samples grown in Oarai and an increase due to some inhomogeneities in the samples can be excluded.

\begin{figure}[h]
\makeatletter
\renewcommand{\thefigure}{S\@arabic\c@figure}
\makeatother
\begin{center}
\includegraphics[width=0.8\columnwidth]{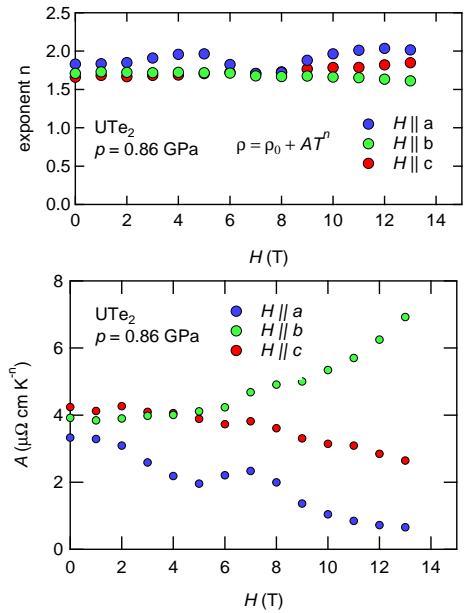}
\caption{(Upper panel) Field dependence of the resistivity exponent $n$ of the temperature dependence of the resistivity, which has been fitted by a power lar $\rho = \rho_0 + AT^n$. At the pressure $p = 0.86$~Gpa we find small deviations from the Fermi liquid $n=2$. (lower panel) Field dependence of the coefficient $A$ of the temperature dependent term of the resistivity. Remakably, $A$ is decreasing for $H \parallel a$ and $c$, while it is increasing for field along the $b$ axis. }
\label{sup_fig4}
\end{center}
\end{figure}

In Fig.~\ref{sup_fig4} we show the analysis of the temperature dependence of the resisitivity for $p = 0.86$ GPa for fields up to 13~T. We want to stress that the magnetoresistiance in the normal state only for $H\parallel b$ axis is positiv, while for the $a$ and $c$ axis it is negativ. The temperature dependence of the resisitivity has been fitted by a power law, $\rho (T) = \rho_0 + AT^n$ in the temperature range between $T_{\rm sc} < T < 6$~K. The field dependence of the exponent $n$ and the coefficient of the temperature dependent term is shown in Fig.~\ref{sup_fig4}. We find that for the resistivity exponent $n \approx 1.7$. It is also possible to force the temeprature dependence to a $T^2$ dependence, however, the fit is significantly less good over this temperature range. In Fig.~\ref{sup_fig4} we show the field dependence of the $A_{\rm n}$ coefficient obtained from the power law.  As expected from the magnetoresistance in the normal state, $A_{\rm n}$ is increasing for field along the $b$ direction, while it is  decreasing for the $a$ and $c$ direction. We also see, that $A_{\rm n} (H)$ for field along the $a$ axis shows some anomaly around 7~T, which may be connected to the Lifshitz transition observed at zero pressure near 6~T. \cite{Niu2020}

\newpage

\section{Resistivity at $p = 1.21$~GPa} 

\begin{figure}[h]
\makeatletter
\renewcommand{\thefigure}{S\@arabic\c@figure}
\makeatother
\begin{center}
\includegraphics[width=0.8\columnwidth]{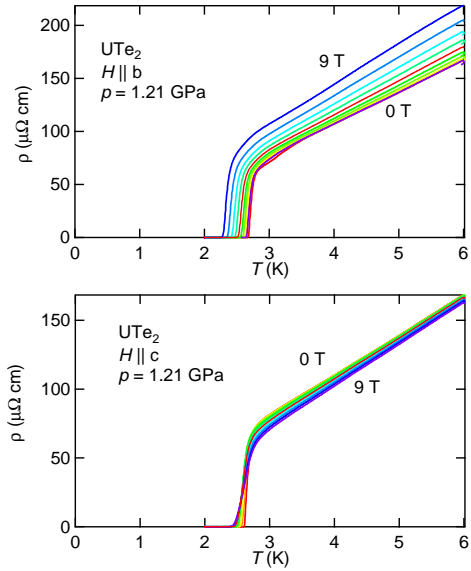}
\caption{(Upper panel) Temperature dependence of the resistivity at 1.21~GPa for $H\parallel b$ (upper panel) and $H\parallel c$ (lower panel).}
\label{sup_fig5}
\end{center}
\end{figure}

In Fig.~\ref{sup_fig5} we show the temperature dependence of the resistivity at $p=1.21$~GPa for $H\parallel b$ (upper panel) and $H\parallel c$ (lower panel) for different fields up to 9~T. The magnetoresistance in the normal state is positive for field along the $b$, and negative for field along the $c$ axis. The temperature dependence is almost linear. Fitting with a power law $\rho(T) = \rho_0 +AT^n$ gives an exponent $n = 1.1$ for both directions in the normal state. It is obvious that the magnetic field suppresses $T_{\rm sc}$ stronger for field applied along the $b$ axis. The onset of superconductivity is almost field independent in the case of $H \parallel c$, in excellent agreement with the results from ac calorimetry from Ref.~\citen{Braithwaite2019}.    

\newpage

\section{Hysteresis of magnetoresistivity at 1.32 GPa}

\begin{figure}[h]
\makeatletter
\renewcommand{\thefigure}{S\@arabic\c@figure}
\makeatother
\begin{center}
\includegraphics[width=0.8\columnwidth]{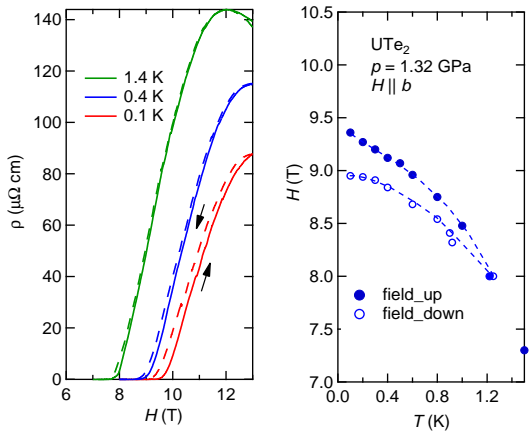}
\caption{(Left panel) Magnetoresistance at $p=1.32$~GPa for different temperatures. Solid lines are field sweep up, dashed lines corresponds to field down. (Right panel) Zoom on the upper cricital field curve indicating the hysteresis between field sweeps up (solid points) and down (down).}
\label{sup_132GPa}
\end{center}
\end{figure}

Near to the critical pressure we observed clear hysteresis in the magnetoresistance as shown in the Fig.~\ref{sup_132GPa} for different temperatures. At low temperature the hysteresis is almost 0.5 T. It vanishes near 1.2~K which corresponds to $\approx 0.5 T_{\rm sc}$. 

\section{$H -T$ phase diagram at 1~GPa}
Here we show magnetoresistivity data for $p = 1$~GPa. These data have been used to drav the phase diagram shown in Fig.~4(a) of the main text. The shown data are exemplary for the $H-T$ phase diagrams shown in Fig.~4(e) of the main text.

\begin{figure}[h]
\makeatletter
\renewcommand{\thefigure}{S\@arabic\c@figure}
\makeatother
\begin{center}
\includegraphics[width=0.8\columnwidth]{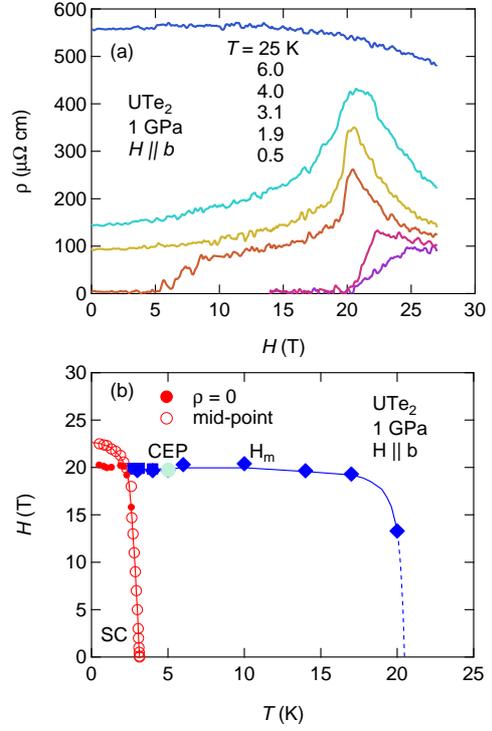}
\caption{(a) Magnetoresistivity at $p = 1$~GPa for different temperature. (b) $H-T$ phase diagram for $p = 1$~GPa. We see that the superconducting transition defined by the $\rho = 0$ criterium coincides with the metamagnetic transition. }
\label{rho_1GPa}
\end{center}
\end{figure}

In Fig.~\ref{rho_1GPa}(a) we show the magnetoresistivity for different temperatures between 0.5~K and 6~K measured in the high field facility in Sendai. At 0.5~K the resistivity is zero up to $H \approx 20$~T. The superconducting transition at 0.5~K is rather broad and only at 25~T the normal state is reached. However, increasing hte temperature the transition get's sharper while the field of zero resistance does not vary up to almost 2~K. For higher temperatures the field of zero resistance decreases, as shown in Fig.~\ref{rho_1GPa}(b) in the phase diagram. From the magnetoresistivity in the normal state we can identify the metamagnetic trnsition field by the almost steplike increase of the resistivity. As shown for $T = 6$~K, above the critical end point the step-like increase disappears and a rather broad maximum defined the cross-over temperature which is connected to the temperature of the maximum of the susceptibility, which is here at 1 GPa near 20~K. 

The important observation is that the metamagnetic transition cuts off the superconductivity defined by the zero resistivity criterium which is nearest to the bulk transition. In Fig. in Fig.~\ref{rho_1GPa}(b) we have plotted the superconducting boundary by both, the zero resistivity and the midpoint of the transition. Clearly, bulk superconductivity seems cut of by the metamagnetic transition.

\section{Analyse of the upper critical field} 

\begin{figure}[h]
\makeatletter
\renewcommand{\thefigure}{S\@arabic\c@figure}
\makeatother
\begin{center}
\includegraphics[width=0.8\columnwidth]{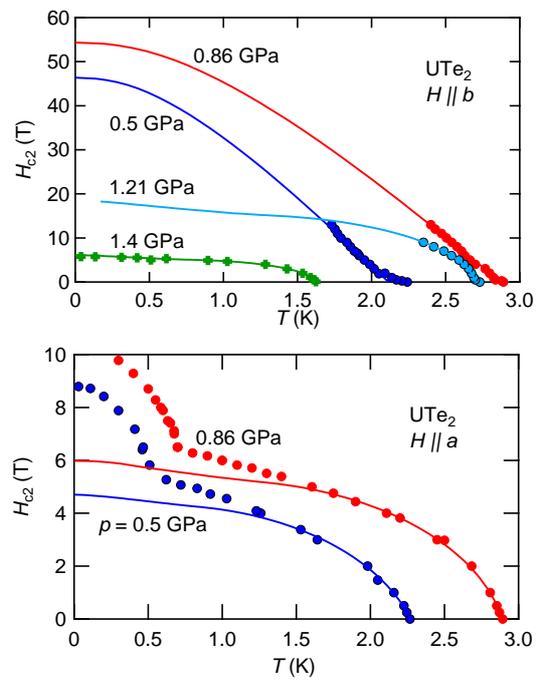}
\caption{(a) Magnetoresistivity at $p = 1$~GPa for different temperature. (b) $H-T$ phase diagram for $p = 1$~GPa. We see that the superconducting transition defined by the $\rho = 0$ criterium coincides with the metamagnetic transition. }
\label{Hc2_fit}
\end{center}
\end{figure}

In Fig.~3 of the main paper we present the pressure dependence of the initial slope $H_{\rm c2}'  = (dH_{\rm c2}/dT)_{T=T_{\rm sc}} $ normalised by  $T_{\rm sc}$ as a function of pressure. In a normal superconductor this allows an estimation of the average Fermi velocity and this of the effective mass of the electrons forming Cooper pairs.In order to get a rather good estimation of the slope, we fitted $H_{\rm c2}$ near $T_{\rm sc}$ taking into account the Pauli limitation and the orbital limitation. It is the orbital limitation which determines the inital slope while the estimation of the Pauli limitation allows for the description of the curvature of $H_{\rm c2}$ near $T_{\rm c}$. Strong coupling effects are neglected. In Fig.~\ref{Hc2_fit} we show the upper critical field for various pressures and the fit of the upper critical field from which we determined the slope and also the $g$ factor shown in Fig.~3(a) of the main article. From this fitting it is obvious that for pressures below 1~GPa for field along the $b$ axis, no Pauli limitation is necessary to reproduce the data. However, for higher pressures, a strong curvature occurs.  Clearly we see that for $H\parallel a$ the Pauli limit can not be neglected. Of course we know that this only a very rough estimation of the behavior and a microscopic model taking into account the correct order parameter, the possible field dependence of the superconducting  pairing and the interplay with the underlying metamagnetic transition has to be taken into account.  

\end{document}